\documentclass[prl, showpacs, preprint]{revtex4}
\usepackage{amsmath}
\usepackage{psfrag}
\usepackage{graphicx}
\usepackage{amssymb}
\usepackage{subfigure}

\newcommand{\Sden}{\rho_{\mathcal{S}}}
\newcommand{\Tden}{\rho_{\mathcal{T}}}
\newcommand{\den}{\rho}

\newcommand{\Ps}{P(\overline{S})}
\newcommand{\Pt}{P(\overline{T})}
\newcommand{\Pst}{P(\overline{S}\,\overline{T})}

\newcommand{\avg}[1]{\left\langle#1\right\rangle}
\newlength{\figwidth}
\DeclareMathOperator{\Li}{Li}

\begin{document}
\title{Shortest path discovery of complex networks}
\author{Attila Fekete}
\email{fekete@complex.elte.hu}
\author{G\'abor Vattay}
\email{vattay@elte.hu}
\affiliation{E\"otv\"os University, Department of Physics of Complex Systems, 
  P\'azm\'any P. s\'et\'any 1/A., H-1117 Budapest, Hungary}

\begin{abstract}
In this Letter we present an analytic study of sampled networks in the case of
some important shortest-path sampling models.  We present analytic formulas for
the probability of edge discovery in the case of an evolving and a static
network model.  We also show that the number of discovered edges in a finite
network scales much more slowly than predicted by earlier mean field models.
Finally, we calculate the degree distribution of sampled networks, and we
demonstrate that they are analogous to a destroyed network obtained by
randomly removing edges from the original network.
\end{abstract}

\maketitle

Complex networks have attracted significant interest in recent years
\cite{AlbertBarabasi02, DorogovtsevMendes02}.  
In most cases, the entire structure of the network is unknown
and one is left with statistical samples of the original network
\cite{LeeKimJeong06,LeskovecFaloutsos06}.  The sampling of Internet topology is
one of the greatest challenges due to its enormous size and decentralized
structure. It motivated numerous studies on the relationship between the original
and the sampled network, including the degree distribution
\cite{LakhinaByersCrovellaXie03,PetermannRios04, ClausetMoore05} and the
expected size of the network \cite{Vigeretal07}. Recently, Internet sampling
methods have emerged that rely on the measurement tool \emph{traceroute}, which
returns the sequence of IP addresses of the network nodes along the path
between the measurement host and a given destination host. 
An abstraction of the network discovery process consists of selecting a set of
source and target nodes and finding the shortest paths between source and
destination pairs. A node or an edge of the network is \emph{discovered} if it
belongs to one of those shortest paths.  
The statistical properties of the discovered network have been studied
extensively by \citeauthor{DallAstaetal05} \cite{DallAstaetal05}.  The
mean-field approximation has been developed in the limit of low source and
target density $\Sden\Tden\ll1$ by neglecting the correlation of different
shortest paths.

In this Letter we present exact results for certain networks.  A surprising new
finding is that the network discovery process is slower in these systems than
it is predicted by the mean-field theory.  While in mean-field approximation
the number of discovered links scales with the product of the number of the
source and target nodes, the new approach predicts a scaling only with their
sum. The lower number of discovered edges is a result of the high degree of
overlapping between shortest paths.  Our other important finding concerns the
degree distribution of the discovered network.  We will show that it is
analogous with a destroyed network where a fraction of the edges of the
original network has been randomly removed. 

We investigate two main discovery strategies. In \emph{peer-to-peer sampling
(P2P)} each node is selected simultaneously for both source and target with
probability $\rho$. Computer applications using the peer-to-peer principle
discover the network this way, hence the name. 
In \emph{disjunct sampling (DI)} each node is selected for source or target but
not for both with probabilities $\Sden$ and $\Tden$. This strategy is used in
Internet mapping projects, where source computers belong to the measurement
infrastructure, while a large number of random addresses are selected as
targets.  
  
We start our analysis with the discovery of a tree. The most important
observation permitting exact calculations in this case is that an edge
separates the tree into two sides. An edge is discovered only if the source and
the target nodes reside on different sides of the edge.  Let us denote the
event that a node is selected as a source or target by $S$ and $T$,
respectively.  Furthermore, we denote the event that at least one source or
target node resides on the 'left' or 'right' side of the edge by $S_{L,R}$ and
$T_{L,R}$, respectively.  The event that a link is discovered, $D$, provided
that its two sides $L$ and $R$ are known, is clearly $D=\left(S_L
T_R\right)+\left(S_R T_L\right)$.  Therefore, we can express the conditional
probability $P(D|L,R)=P(S_L|L,R)\,P(T_R|L,R)+P(S_R|L,R)\,P(T_L|L,R)-P(S_L
T_L|L,R)\,P(S_R T_R|L,R).$  The probabilities arising in this expression can be
calculated easily: $P(S_{\lambda}\mid L,R)=1-P^{N_{\lambda}}(\overline{S})$,
$P(T_{\lambda}\mid L,R)=1-P^{N_{\lambda}}(\overline{T})$ and
$P(S_{\lambda}T_{\lambda}\mid
L,R)=1-P^{N_{\lambda}}(\overline{S})-P^{N_{\lambda}}(\overline{T})
+P^{N_{\lambda}}(\overline{S}\,\overline{T})$, 
where $\lambda=L$ or $R$, $N_L$ and $N_R$ are the number of nodes on the two sides of
the link, and the overlines denote complement events.

Let us consider an evolving network where one new edge is attached randomly to
the nodes of the existing network.  The structure of this network will be a
tree.  Since the network is connected the cluster sizes $N_L$ and $N_R$ must
satisfy the relation $N_L+N_R=N$, where $N$ is the size of the whole network.
In the  thermodynamic limit $N\to\infty$ 
we obtain $P(D\mid N_L)=1-{\sigma}^{N_L}$, where we have introduced
$\sigma=\Pst$.  
\renewcommand{\Pst}{\sigma}
The probability $\Pst$ in the different sampling
models is related to the source and target densities in a simple way:
\begin{equation}
  \Pst=
  \begin{cases}
    1-\rho &\text{P2P}\\
    1-\Sden-\Tden&\text{DI},
  \end{cases}
  \label{eq:sampling_cases}
\end{equation}
where $\den,\Sden,\Tden\in\left[0,1\right]$, $\Sden+\Tden\le1$. 
If $\Sden+\Tden\ll1$ in the DI sampling model, then we can 
write $P(D\mid L)\approx1-\exp\left(-\frac{\Sden+\Tden}{N}b_e\right)$,
where $b_e=N_L\left(N-N_L\right)$ is the number of shortest paths
that traverse a given link, called \emph{betweenness centrality}.  
Compare this result with the mean field model of 
\citeauthor{DallAstaetal05} \cite{DallAstaetal05}: 
$P(D_{\mathrm{m.f.}}\mid b_e)\approx1-\exp\left(-\Sden\Tden b_e\right)$.

The probability of finding an arbitrary edge by trace\-route probes
can be given now straightforwardly:
\begin{equation}
  \pi_d = \sum_{N_L=0}^{\infty}P(D\mid N_L)P(N_L)=1-H_1(\Pst),
  \label{eq:disc_prob_evolving}
\end{equation}
where $H_1(z)=\sum_{N_L} P(N_L)z^{N_L}$ is the generating function of the 
cluster size distribution $P(N_L)$.

\begin{figure}[tb]
  \psfrag{p}[c][c][1.2]{$\den$}
  \psfrag{P}[c][c][1.2]{$\pi_d(\den)$}
  \psfrag{n}[c][c][1.2]{$n$}
  \psfrag{n_d}[c][c][1.2]{$n_d$}
  \psfrag{a=0}[r][r]{$a=0$}
  \psfrag{a=1}[r][r]{$a=1$}
  \psfrag{a=inf}[r][r]{$a=\infty$}
  \begin{center}
    \resizebox{\figwidth}{!}{\includegraphics{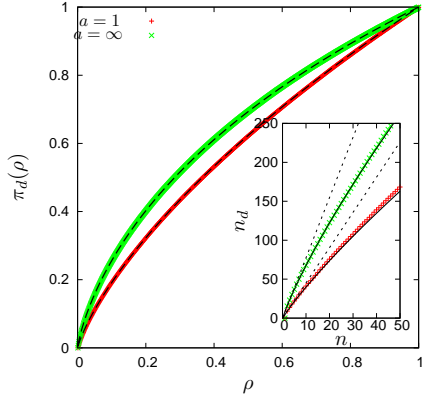}}
  \end{center}
  \caption{(Color online) Discovery probability of edges $\pi_d(\den)$ as the 
  function of the measurement node density $\rho$ in P2P sampling of evolving 
  trees.  Data points are averaged over $100$ realizations of $N=10000$ node 
  random trees with $a=1$ and $+\infty$.  Dashed lines show the analytic 
  solution (\ref{eq:disc_prob_spec}) with $\Pst=1-\den$.  The inset shows 
  the expected number of discovered edges $\avg{n_d}$ as the function 
  of the number of the measurement nodes $n\ll N$.  The solid line
  represents (\ref{eq:disc_nodes_dominating}) for P2P sampling, whereas the 
  dotted line shows its leading term $\avg{l}n/2$ with $\avg{l}=9.045$ and 
  $15.48$ for $a=1$ and $+\infty$, respectively.}
  \label{fig:disc_prob_AS}
\end{figure}
Expression (\ref{eq:disc_prob_evolving}) has been tested on the
Dorogovtsev--Mendez (DM) network growth model~\cite{DorogovtsevMendesSamukhin00}, a
generalization of the Barab\'asi--Albert (BA) model~\cite{BarabasiAlbert99}, where
new nodes with $m$ new links are attached to old nodes with degree 
dependent probability $\Pi(k_i)=\frac{k_i-m+am}{\sum_i\left(k_i-m+am\right)}$,
where $a\ge0$.  The growing tree corresponds to $m=1$.  
We calculated the distribution $P(N_L)$ for this model analytically in
Ref.~\cite{FeketeVattayKocarev06b}.  The generating function can be expressed
in terms of hypergeometric functions $H_1(z)=z\,_2F_1(1-\alpha, 1, 2-\alpha;
z)-z\frac{1-\alpha}{2-\alpha}\,_2F_1(2-\alpha, 1, 3-\alpha; z)$ and
$\alpha=\frac1{1+a}$.  At $a=1$ we recover the original BA
preferential attachment model with scale-free degree distribution
and at $a=+\infty$ we obtain uniform attachment probability 
with exponential degree distribution. In these cases $\pi_d$
can be expressed with elementary functions
\begin{equation}
  \pi_d=
  \begin{cases}
    -\frac{1-\Pst}{\Pst}\ln\left(1-\Pst\right) &\text{if $a=+\infty$ (i.e. $\alpha=0$)},\\
    \frac{1-\Pst}{2\sqrt{\Pst}}\ln\frac{1+\sqrt{\Pst}}{1-\sqrt{\Pst}} &\text{if $a=1$ (i.e. $\alpha=1/2$)}.
  \end{cases}
  \label{eq:disc_prob_spec}
\end{equation}

Figure~\ref{fig:disc_prob_AS} shows simulations for the P2P sampling
model at $\alpha=0$ and $1/2$.  The analytic results
(\ref{eq:disc_prob_spec}), plotted with dashed lines, fit the simulation data
excellently.  

\begin{figure}[tb]
  \psfrag{p}[c][c][1.2]{$\rho$}
  \psfrag{P}[c][c][1.2]{$P(e)$}
  \psfrag{z=0.5}[r][r]{$\avg{k}=0.5$}
  \psfrag{z=1.0}[r][r]{$\avg{k}=1$}
  \psfrag{z=2.0}[r][r]{$\avg{k}=2$}
  \psfrag{z=4.0}[r][r]{$\avg{k}=4$}
  \begin{center}
    \resizebox{\figwidth}{!}{\includegraphics{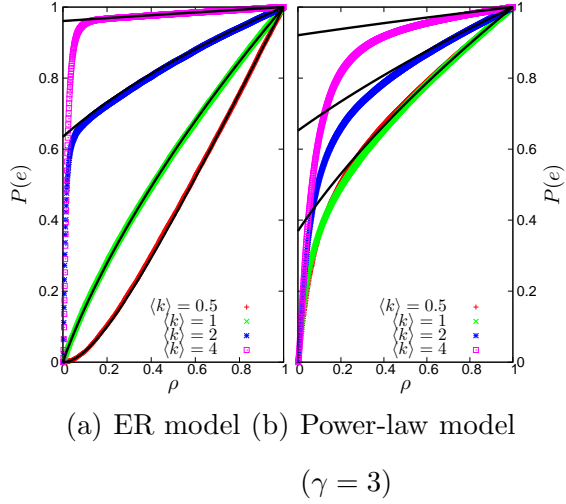}}\label{subfig:disc_prob_ER}\\
  \small
  \begin{tabular}{rlrl}
  (a) & ER model & (b) & Power-law model\\
      &          &     &($\gamma=3$)
  \end{tabular}
  \end{center}
  \caption{(Color online) Discovery probability of edges as the function 
  of the fraction of the measurement nodes $\rho$ in static networks. $100$ all-to-all samplings
  were averaged in $N=10000$ size networks with average degrees $\avg{k}=0.5,1,2$ and $4$.  
  Solid lines show the analytic formula (\ref{eq:disc_prob_hidden_AS}).}
  \label{fig:disc_prob_HIDDEN}
\end{figure}

From the point of view of the efficiency of the discovery process, it is
important to calculate how many edges can be discovered with a given number of
source $n_S$ and target nodes $n_T$.  For the Internet discovery the disjunct
sampling model is relevant, where $\varrho_T+\varrho_S=(n_T+n_S)/N=n/N=1-\Pst\ll1$.
The series expansion of (\ref{eq:disc_prob_evolving}) yields 
$\pi_d=1-\sum_{N_L} P(N_L)\left(1-\frac{n}{N}\right)^{N_L}$. We can rearrange the 
series by adding and subtrating the terms $1-n\frac{N_L}{N}$ and averaging them
separately
$\pi_d=\frac{n\langle N_L\rangle}{N}-\sum_{N_L} P(N_L)\left[\left(1-\frac{n}{N}\right)^{N_L}-1+n\frac{N_L}{N}\right].$

Several authors have pointed out that the distribution of $b_e=N_L\left(N-N_L\right)$ 
follows a universal power-law tail in trees with exponent $-2$ \cite{SzaboAlavaKertesz02, 
KimNohJeong04, FeketeVattayKocarev06b}.  It also implies that
asymptotically $P(N_L)\approx c N_L^{-2}$ in an arbitrary tree for $N_L\gg1$.
Specifically, $c=1-\alpha$ in the DM model.  Using this asymptotic form we can calculate
the leading behaviour in the $N\rightarrow\infty$ limit
$\pi_d=\frac{n\langle N_L\rangle}{N} -c \Li_2(1-n/N)+c\frac{\pi^2}{6}-c
\frac{n}{N}(\ln N-\gamma)$, where $\Li_2(x)$ is the dilogarithm function and
$\gamma\approx0.5772$ is the Euler constant.  For small argument
$\Li_2\left(1-x\right)$ can be expanded by using Euler's reflection formula
$\Li_2 \left(1-x \right)= -\Li_2 \left(x \right)+
\frac{\pi^2}{6}-\ln(x)\ln(1-x)\approx -x + \frac{\pi^2}{6}+x\ln(x)+\dots$.
Finally we get $\pi_d=\frac{n\langle N_L\rangle}{N}
+c\frac{n}{N}-c\frac{n}{N}\ln n-c \frac{n}{N}\gamma$.

To process this further, let us express the term $\avg{N_L}$ more straightforwardly.  
The sum of $b_e$ for all edges clearly equals the total length of the shortest 
paths between all possible pairings of nodes:
$\sum_{e\in E}b_e=\sum_{i,j\in V}l_{i,j}$.  Since
$\avg{b}=\frac1{N-1}\sum_{e\in E}b_e$ and
$\avg{l}=\frac2{N(N-1)}\sum_{i,j\in V}l_{i,j}$ we can write
$\avg{l}N/2=\avg{b}$.  Therefore, the average branch size can be given as
$\avg{N_L}=\avg{l}/2+\avg{N_L^2}/N$, where $\avg{N_L^2}/N=\frac1{N}\sum_{N_L=1}^{N}\frac{c}{N_L^2}N_L^2=c$.
For a large, but finite network the average number of discovered edges is
$\avg{n_d}=\left(N-1\right)\pi_d$, that is
\begin{align}
  \avg{n_d}&\approx n\left(\frac{\avg{l}}{2}-c\ln n+2c-c\gamma\right)
  \label{eq:disc_nodes_dominating}
\end{align}
in the limit $1\ll n=n_S+n_T\ll N$,  The above result shows that $\avg{n_d}$ depends on the sum of $n_S$
and $n_T$.  This is in contrast to the mean field model, which predicts that
$\avg{n_d}$ scales with the product of $n_S$ and $n_T$.  The logarithmic term
of (\ref{eq:disc_nodes_dominating}) accounts for the possibility that a new
measurement node is placed at a node discovered by previous measurement nodes.
The inset of Fig.~\ref{fig:disc_prob_AS} displays simulation results and the
formula corresponding to the P2P sampling.

\begin{figure}[tb]
  \psfrag{v}[c][c][2]{$v$}
  \psfrag{N1}[c][c][2]{$N_1$}
  \psfrag{N2}[c][c][2]{$N_2$}
  \psfrag{N3}[c][c][2]{$N_3$}
  \psfrag{N4}[c][c][2]{$N_4$}
  \psfrag{N5}[c][c][2]{$N_5$}
  \psfrag{N6}[c][c][2]{$N_6$}
  \psfrag{Nk}[c][c][2]{$N_{k}$}
  \begin{center}
    \resizebox{0.7\figwidth}{!}{\includegraphics{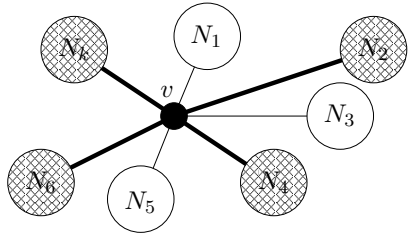}}
  \end{center}
  \caption{(Color online) Schematic diagram of an arbitrary vertex $v$ with degree 
$k$ and the emerging branches with sizes $N_1, N_2, \dots, N_k$.  Shaded circles
  represent branches where measurement nodes can be found in.  Thick lines symbolize 
  the discovered edges of node $v$.}
  \label{fig:schema_degree_static}
\end{figure}

We continue with the analysis of a static model where nodes are randomly
connected with a prescribed degree distribution $p_k$. This 'configuration
model' is a generalization of the Erd\H os-R\'enyi (ER) model
\cite{ErdosRenyi60}, where the degree distribution is Poissonian.  It has been
shown in \cite{NewmanStrogatzWatts01} that the generating function of branch
sizes $H_1(z)$ satisfies the implicit equation $H_1(z)=zG'_0(H_1(z))/\avg{k}$, where
$G_0(z)=\sum_k p_k z^k$ is the generating function of the degree distribution.
In the configuration model loops become irrelevant in the thermodynamic limit
$N\to +\infty$ and each edge is a part of a tree. Here, $N_L$ and $N_R$ are
independent and the joint probability function has a product form
$P(N_L,N_R)=P(N_L)P(N_R)$.  The summation in $\pi_d$ can be carried out
separately for $N_L$ and $N_R$, which yields
\renewcommand{\Pst}{P(\overline{S}\,\overline{T})}
\begin{align}
  \pi_d&=2\left(1-H_1(\Ps)\right)\left(1-H_1(\Pt)\right)\notag\\
  &-\left(1-H_1(\Ps)-H_1(\Pt)+H_1(\Pst)\right)^2.
\end{align}
In the case of P2P discovery this can be reduced to
\begin{equation}
  \pi_d=\left(1-H_1(1-\den)\right)^2.
  \label{eq:disc_prob_hidden_AS}
\end{equation}

This formula can be tested on the ER model, with 
$G_0(z)=e^{<k>(z-1)}$. The cluster size distribution can be given by the
Lambert W-function $H_1(z)=-W(-\avg{k} e^{-\avg{k}} z)/\avg{k}$.  Simulation
results are presented in Fig.~\ref{fig:disc_prob_HIDDEN}(a). 
The analytic result (\ref{eq:disc_prob_hidden_AS}) is also shown for
comparison.  One can see that it 
is discontinuous at zero
density if $\avg{k}>1$, when a giant component emerges in the network. The
simulation data deviates from the analytic solution around the discontinuity
due to finite-scale effects.   The size of the jump is
$P_0=\left(1-H_1(1)\right)^2$, which is precisely 
the probability of infinitely large branches being attached to both sides of an
edge.  If $P_0$ is regarded as an order parameter, the observed phenomenon
resembles a phase transition at $\avg{k}=k_c=1$.

We also generated networks with power-law degree distribution using the
hidden-variable model introduced in \cite{GohKahngKim01,
CaldarelliCapocciRiosMunoz02, Soderberg02, BogunaPastor03}. Simulations are
shown in Fig.~\ref{fig:disc_prob_HIDDEN}(b) with degree exponent $\gamma=3$.
Note that the analytic solution is discontinuous at zero density, i.e.
$P_0>0$, for all $\avg{k}>0$.  The phase transition can be observed again,
since the analytic solution---and $P_0$---is independent of $\avg{k}$ below a
critical point $k_c(\gamma)=\frac{\zeta(\gamma-1)}{\zeta(\gamma)}$.  Indeed,
data points almost collapse at $\avg{k}=0.5$ and $1$ which are below
$k_c(\gamma=3)\approx1.3684$.    The phenomenon occurs when the degree generating
function $G'_0(z)$ depends linearly on $\avg{k}$.  This is characteristic of pure
power-law distributions until $\avg{k}$ is below the critical value $k_c$.

\newcommand{\ks}{k'}
\newcommand{\kd}{k'}
\newcommand{\Vd}{V_{\mathrm{d}}}
\begin{figure}[tb]
  \psfrag{p_k}[c][c][1.2]{$P(\ks)$}
  \psfrag{p_q}[c][c][1.2]{$P(q')$}
  \psfrag{p}[c][c][1.2]{$\den$}
  \psfrag{k=2}[r][r]{$\ks=2$}
  \psfrag{k=3}[r][r]{$\ks=3$}
  \psfrag{k=4}[r][r]{$\ks=4$}
  \psfrag{k=5}[r][r]{$\ks=5$}
  \psfrag{k=6}[r][r]{$\ks=6$}
  \psfrag{k=7}[r][r]{$\ks=7$}
  \psfrag{q=1}[r][r]{$q'=1$}
  \psfrag{q=2}[r][r]{$q'=2$}
  \psfrag{q=3}[r][r]{$q'=3$}
  \psfrag{q=4}[r][r]{$q'=4$}
  \psfrag{q=5}[r][r]{$q'=5$}
  \psfrag{q=6}[r][r]{$q'=6$}
  \begin{center}
    \resizebox{\figwidth}{!}{\includegraphics{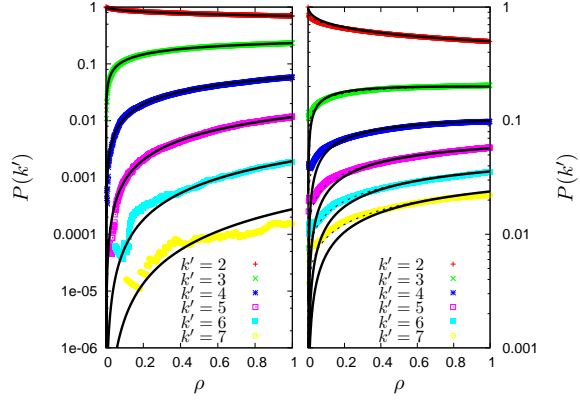}}\\
  \small
  \begin{tabular}{rlrl}
  (a) & Static ER network & (b) & Evolving BA tree with\\
      & with $\avg{k}=1$  &     & degree exponent $\gamma=3$.
  \end{tabular}
  \end{center}
  \caption{(Color online) The probability of discovered degree $P_{\mathrm{d}}(\ks)$ as the 
  function of $\den$ in P2P sampling model for $\ks=2,3,\dots,7$.  The original
  networks are $N=10^4$ node graphs.  Data points are averaged for $10$ 
  networks with $10$ samplings in each realization.  Solid lines consist of
  analytic solution (\ref{eq:deg_dist_disc}) for (a) ER and (b) 
  BA network models, respectively.  Exact solution for the evolving BA model 
  is shown with dotted lines for comparison.}
  \label{fig:degree_prob}
\end{figure}

\renewcommand{\Pst}{\sigma}
Now we turn our attention to the degree distribution $P_{\mathrm{d}}(\ks)$ of
the discovered nodes.  In our analysis we consider only the contribution of
those shortest-paths to $\ks$ which \emph{traverse} a given node.  We will show
that $P_{\mathrm{d}}(\ks)$ is analogous to the degree distribution of a
partially severed network obtained by random edge pruning.  This duality between the
sampling and the destruction of networks is very surprising considering the
striking differences between the two processes.

Let us consider a node $v$ with original degree $k$.  If every link is removed
independently with probability $p$, then $\kd$, the degree of the
node after random edge removal, will follow a binomial distribution:
$P(\kd\mid k)=\binom{k}{\kd}\left(1-p\right)^{\kd}p^{k-\kd}$.
Consequently, 
\begin{equation}
  P_{\mathrm{pruned}}(\kd)
  =\sum_{k=\kd}^{\infty}\binom{k}{\kd}\left(1-p\right)^{\kd}p^{k-\kd}P_0(k).
  \label{eq:deg_dist_dest}
\end{equation}

Regarding the sampling process we examine a randomly selected node of the
discovered network $v\in\Vd$ in the static model first.  Let us suppose that
the sizes of the branches with original degree $k$ are $N_1, N_2, \dots, N_{k}$
(see Fig.~\ref{fig:schema_degree_static}).  For the sake of simplicity we
discuss only the P2P sampling model, where the probability of placing a
measurement node in branch $i$ is simply $\left(1-\Pst^{N_i}\right)$.  Since
branch sizes are independent we can average over $N_i$ separately.  The results we obtain indicate 
that measurement nodes can be found in different branches with probability
$1-H_1(\Pst)$.

We can see from Fig.~\ref{fig:schema_degree_static} that the degree of a 
discovered node $\ks$ equals the number of branches where measurement nodes 
can be found in.  
It follows that $P_{\mathrm{d}}(\ks\mid k)
=\frac1{P(v\in\Vd\mid k)}\binom{k}{\ks}\left(1-H_1(\Pst)\right)^{\ks}H_1^{k-\ks}(\Pst)$,
where $2\le\ks\le k$.  The subscript of $P_{\mathrm{d}}$ refers to the 
probability distribution restricted to the discovered network.
In order to obtain the distribution of $\ks$ one should average this
probability over $P_{\mathrm{d}}(k)$, the distribution of \emph{the original 
degrees of the discovered nodes}.  This distribution can be obtained by
$P_{\mathrm{d}}(k)=\frac{P(v\in\Vd\mid k)P_0(k)}{P(v\in\Vd)}$, so
\begin{equation}
  P_{\mathrm{d}}(\ks)
  =\frac{\sum_{k=\ks}^{\infty}
  \binom{k}{\ks}\left(1-H_1(\Pst)\right)^{\ks}H_1^{k-\ks}(\Pst)P_0(k)}{P(v\in\Vd)},
  \label{eq:deg_dist_disc}
\end{equation}
where
$\ks\ge2$ and $P(v\in\Vd)
  =1-G_0(H_1(\Pst))-\left(1-H_1(\Pst)\right)G_0'(H_1(\Pst))$
It is evident from (\ref{eq:deg_dist_dest}) and
(\ref{eq:deg_dist_disc}) that $P_{\mathrm{d}}(\ks)$ equals
$P_{\mathrm{pruned}}(\kd)$---normalized properly for $\kd\ge2$---if
$p=H_1(\Pst)$.  In other words the discovered network is equivalent with an edge
destroyed one.

\renewcommand{\ks}{q_{\mathrm{d}}}
In the case of an evolving network at least one of the branches, say $N_k$,
tends to infinity as $N\to\infty$, so the probability that a
measurement node can be found in the $k$th branch tends to 1.  In order to
circumvent this effect let us redefine the network in such a way that every
link should be directed toward the gigantic side of the network.  Let $q=k-1$ 
denote the in-degree of nodes in this directed network.  It is easy to
see that the discovered \emph{in-degree} $\ks$ will be equal to the number of
branches where measurement nodes can be found in.
We can follow the same procedure as in the case of the static model.  We only
need to replace $k_{\textrm{d}}$ and $k$ in (\ref{eq:deg_dist_disc}) with the
corresponding in-degrees $\ks$ and $q$, and the normalization constant with
$P(v\in\Vd)=1-G_0^{(\mathrm{in})}(H_1(\Pst))$.

Simulation results are shown for both static and evolving networks in
Fig.~\ref{fig:degree_prob}.  Note that we have assumed above that $H_1(\Pst)$
is independent of $q$.  This is only an approximation in the case of the evolving
network model.  However, $H_1(\Pst\mid q)$ can be calculated exactly for the DM
model, which is shown with dotted lines \cite{FeketeVattay}.

In conclusion we presented a study of network discovery processes.  We 
derived analytically the probability of founding an arbitrary link of the
network via shortest-path network discovery.  We considered both static and 
evolving random netwoks with various sampling scenarios.  We also demonstrated
an important duality between the discovery of networks by shortest paths
and the destruction of the same network by edge removal.

\end{document}